\documentclass[showpacs,twocolumn,pra,superscriptaddress,notitlepage]{revtex4-2}

\usepackage{graphicx}
\usepackage{rotating}
\usepackage{color}
\usepackage{amsmath}
\usepackage{siunitx}
\usepackage[ruled]{algorithm2e}
\begin{document}

%%%%%%%%%%%%%%%%%%%%%%%%%%%%%%%%%%%%%%%%%%%%%%%%%%%%%%%%%%%%%%%%%%
%                      Title and Authors                         %
%%%%%%%%%%%%%%%%%%%%%%%%%%%%%%%%%%%%%%%%%%%%%%%%%%%%%%%%%%%%%%%%%%
\title{Divide-and-conquer variational quantum algorithms for large-scale electronic structure simulations}

\author{Huan Ma}
\affiliation{Hefei National Laboratory, University of Science and Technology of China, Hefei 230088, China}

\author{Yi Fan}
\affiliation{Hefei National Research Center for Physical Sciences at the Microscale, University of Science and Technology of China, Hefei 230026, China}

\author{Jie Liu}
\email{liujie86@ustc.edu.cn}
\affiliation{Hefei National Laboratory, University of Science and Technology of China, Hefei 230088, China}

\author{Honghui Shang}
\email{shanghonghui@ict.ac.cn}
\affiliation{State Key Laboratory of Computer Architecture, Institute of Computing Technology, Chinese Academy of Sciences, Beijing, 100190, China}
  
\author{Zhenyu Li}
\affiliation{Hefei National Laboratory, University of Science and Technology of China, Hefei 230088, China}
\affiliation{Hefei National Research Center for Physical Sciences at the Microscale, University of Science and Technology of China, Hefei 230026, China}

\author{Jinlong Yang}
\email{jlyang@ustc.edu.cn}
\affiliation{Hefei National Laboratory, University of Science and Technology of China, Hefei 230088, China}
\affiliation{Hefei National Research Center for Physical Sciences at the Microscale, University of Science and Technology of China, Hefei 230026, China}

\date{\today}% It is always \today, today,

%%%%%%%%%%%%%%%%%%%%%%%%%%%%%%%%%%%%%%%%%%%%%%%%%%%%%%%%%%%%%%%%%%
%                            Abstract                            %
%%%%%%%%%%%%%%%%%%%%%%%%%%%%%%%%%%%%%%%%%%%%%%%%%%%%%%%%%%%%%%%%%%
\begin{abstract}
%Quantum computing grows rapidly in resent years, and it has the potential application for material and biological systems. However, in the near-term quantum device, the number of qubits has been limited, so the simulation of the large realistic system is still challenging. Here in this work, a problem decomposition method has been adopted to use small number of qubits for a large molecule. Then we integrate the adaptive VQE method with the decomposition method, and we can reduce the simulation size and shallow the quantum circuit. In this way, the realistic chemical problem can be solved on near-term quantum computers. 
Exploring the potential application of quantum computers in material design and drug discovery has attracted a lot of interest in the age of quantum computing. However, the quantum resource requirement for solving practical electronic structure problems are far beyond the capacity of near-term quantum devices. In this work, we integrate the divide-and-conquer (DC) approaches into the variational quantum eigensolver (VQE) for large-scale quantum computational chemistry simulations. Two popular divide-and-conquer schemes, including many-body expansion~(MBE) fragmentation theory and density matrix embedding theory~(DMET), are employed to divide complicated problems into many small parts that are easy to implement on near-term quantum computers. Pilot applications of these methods to systems consisting of tens of atoms are performed with adaptive VQE algorithms. This work should encourage further studies of using the philosophy of DC to solve electronic structure problems on quantum computers.          
\end{abstract}

\maketitle
\section{Introduction}
Electronic structure simulations are essential in computational-aided material design and drug discovery, which are two anticipated applications of quantum computing. Many recent efforts have been devoted to the development of advanced quantum algorithms for the exact solution of the electronic structure (ES) on noisy intermediate-scale quantum (NISQ) devices.~\cite{AspDutLov05,DuXuPen10,PerMcCSha14,MalBabKiv16,KanMezTem17,HemMaiRom18,AruAryBab20,Pre18,CaoRomOls19,McAEndAsp20,ZhoWanDen20,LiuFanLi22} The variational quantum eigensolver (VQE) is one of leading algorithms to implement ES simulations on near-term quantum computers.~\cite{PerMcCSha14,McCRomBab16,RomBabMcC18,AruAryBab20} Due to the variational nature and relatively shallow quantum circuit compared to the quantum phase estimation, the VQE mitigates the effect of noise in quantum computing. In particular, the combination of different error mitigation techniques enabled 12-qubit Hartree-Fock VQE calculations (with 72 two-qubit gates) on a superconductor quantum computer.~\cite{AruAryBab20} Although many exciting VQE experiments have been recently reported for ES simulations,~\cite{KanMezTem17,ParHohMcM19,MaGovGal20,AruAryBab20,OllKanChe20,KawLloCoo21} solving industrially related problems using the VQE is still a great challenging work due to the error accumulation. 
%this is a classically tractable problem, it paves the way for large-scale quantum computational chemistry simulations. While, because only very shallow quantum circuits can be efficiently implemented on near-term quantum devices, an exact solution of the electronic structure using the VQE is still a great challenging work.
%to find an exact solution of the electronic structure requires the implementation of the state-of-the-art wave function ansatzes, such as unitary coupled cluster~\cite{Eva11,HarShiScu18,PerMcCSha14} or adaptive derivative-assembled pseudotrotter (ADAPT),~\cite{GriEcoBar19} in the VQE. These ansatzes result in deep quantum circuits that goes beyond the capacity of current quantum devices even for few-qubit VQE calculations due to the error accumulation. 

%The electronic structure method is an essential way to get a better understanding of the material, chemistry and biological systems. The full configuration interaction (full CI) method can give the exact solution of the electronic structure, but it grow exponentially with the system size, which can not be used into the realistic systems. In order to solve this problem, quantum computing is arising, because it can overcome the exponential scaling in quantum chemistry simulation. The variational quantum eigensolver (VQE) is such a quantum- classical algorithms which designed for the noisy intermediate- scale quantum (NISQ).

To apply the VQE in ES problems involving tens of qubits, which may find use in practical applications, one can resort to fault-tolerant quantum computers scaling up to tens or even hundreds of qubits. This represents a long-term plan of quantum computing with quantum error correction. As a potential strategy tailored for NISQ devices, the philosophy of divide-and-conquer (DC) has been recently introduced in quantum computing.~\cite{BauWecMil16,KreGarLam16,YamMatNar18,KawLloCoo21,LiHuaCao21} A complicated ES problem can be divided into as many parts as possible according to available qubit counts, gate fidelity and coherence time. A general framework of combining the VQE with the DC strategy has been proposed for large-scale quantum simulations of the ES.~\cite{YamMatNar18} From a quantum computational chemistry perspective, there are two key attractive aspects of the DC approaches:
\begin{enumerate}
\item The size of subsystems that determines the number of qubits used in quantum computing can adapt to available quantum resources. And then massively parallel quantum simulations of subsystems can be performed on NISQ devices. Finally, the ES of subsystems is combined to obtain the ES of the full system. This maximally utilizes small-scale quantum computational resources in practical applications.  
\item The circuit depth in each subsystem simulation is significantly reduced in comparison with that in the simulation of the full system. For example, the circuit depth of the UCC with single and double excitation (UCCSD) scales as $\mathcal{O}(N^4)$ with $N$ being the system size. If the full system is divided into $N_s$ subsystems, the circuit depth of UCCSD for each subsystem is in principle reduced by $N_s^4$. The low-depth circuits can be effectively combined with error mitigation techniques to reduce the total error rate.   
\end{enumerate}
In the age of quantum computing without error correction, it is appealing to explore these two aspects by combining the VQE with the DC techniques to enable materials simulations on NISQ devices.   

Two popular DC schemes for large-scale ES simulations are fragment-based approaches~\cite{AkiPre15} and quantum embedding theories.~\cite{SunCha16} Fragment-based approaches usually partition a system according to chemical information, such as chemical functional groups and chemical bonds. A variety of fragment-based approaches, such as the fragment molecular-orbital method,~\cite{FedNagKit12} the molecular fragmentation with conjugated caps method,~\cite{ZhaZha03,WanLiuZha13} and the (generalized) many-body expansion,~\cite{DahTru07,RicHer12,LiuHer16} have been proposed for simulations. The accuracy of fragment-based approaches depends on the molecular fragmentation schemes and posteriori corrections to many-body and long-range interactions.~\cite{AkiPre15} In contrast, quantum embedding theories provide a more flexible way to partition the large systems. The interaction between the subsystem and the environment is described by including the bath orbitals in the subsystem calculations. A pilot of implementation the density matrix embedding theory (DMET) within the framework of the VQE has been recently realized for a ring of 10 hydrogen atoms on a trapped-ion quantum computer.~\cite{KawLloCoo21} A numerical simulation of C$_{18}$ with the cc-pVDZ basis (up to 144 qubits) has been performed with the DMET-VQE method.~\cite{LiHuaCao21} Although these preliminary works have demonstrated the advantage of the DC strategy in quantum computing, the application of the DC approaches within the VQE framework to practical ES problems is still an open question.

In this work, we integrate the DC approaches into adaptive VQE algorithms to explore the ES of complex systems. To enable quantum simulations of systems consisting of tens of atoms, a large system is first partitioned into many fragments. The many-body expansion (MBE) fragmentation approach is used to describe high-order corrections, namely interactions among different fragments.  To further pursuit shallow quantum circuits, an adaptive VQE algorithm that iteratively builds a compact wave function ansatz~\cite{GriEcoBar19} is introduced to find an exact solution of the ES in subsystems. The adaptive VQE algorithm is also integrated with the DMET approach for comparison. The MBE-VQE method is able to provide a reasonable character of configuration stability. In Section~\ref{sec:method}, we give a brief introduction of the DC approaches, including the MBE and DMET, and adaptive VQE algorithms. In Section~\ref{sec:result}, we perform benchmark tests with the hydrogen chain and C$_{18}$ ring for MBE-VQE and then apply this method to study the relative energies of water hexamers.

\section{Method}
\label{sec:method}
Given a set of Hartree-Fock (HF) orbitals, the second-quantized Hamiltonian can be written as:
	\begin{equation}
		\hat{H}=\sum_{p,q}{h_{q}^{p}a_{p}^{\dagger}a_{q}}
		+\frac{1}{2}{\sum^N_{\substack{p,q\\r,s}}} {v_{sr}^{pq} a_{p}^{\dagger}a_{q}^{\dagger}a_{r}a_{s}}
		\label{eq-ham-pbc}
	\end{equation}
where $h_{q}^{p}$ and $v_{sr}^{pq}$ are one- and two-electron integrals in molecular orbital basis, respectively. In a quantum simulation, the creation and annihilation operators $\{{a_{p}^{\dagger}}, a_{q}\}$ are mapped to qubit operators using the Jordan-Wigner or Bravyi-Kitaev transformation. The energy can be obtained by measuring and then summing up the expectation values of Pauli strings $\{\hat{P}\}\in\{I, \sigma_{x},\sigma_{y},\sigma_{z}\}^{\otimes n} $ as
	\begin{equation}
	    E=\langle \Psi(\theta)|\hat{H}|\Psi(\theta) \rangle
		\label{eq:qubit-ham}=\sum_\mu C_\mu \langle \hat{P}_\mu \rangle
	\end{equation}
For large-scale simulations, the number of qubits and the circuit depth become unaffordable on near-term quantum devices. In order to reduce the usage of the computational resources, it is necessary to resort to the DC strategy. In the following, we will introduce two popular DC approaches.  

%The ground-state wave function and energy are obtained by solving the stationary Schr\"odinger equation:
%	\begin{equation}
%		\hat{H}|\Psi\rangle=E_{0}|\Psi\rangle
%	\end{equation}

\subsection{Divide-and-Conquer methods}

\subsubsection{Many Body Expansion Theory}
The many-body expansion theory presents a simple and intuitive fragmentation scheme for large-scale simulations. Here, a large system is first partitioned into many nonoverlapping fragments, often called monomers, and then the many-body effect is captured by high-order corrections from multiple fragments, namely dimers, trimers, and so forth. Without any approximation, the total energy of the whole system can be written as
\begin{equation} \label{eq:MBE}
E=\sum_{i}^{n} E_{i}+\sum_{i<j}^{C^n_2} E^{(2)}_{i,j}+\sum_{i<j<k}^{C^n_3}  E^{(3)}_{i,j,k}+\cdots
\end{equation}
Where $E_i$ represents the energy of $i$-th fragment and $E^{(n)}$ are $n$-order energy corrections defined as
\begin{equation}
\begin{gathered}
E^{(2)}_{i,j}=E_{i,j}-E_{i}-E_{i} \\
E^{(3)}_{i,j,k}=E_{i,j,k}-E^{(2)}_{i,j}-E^{(2)}_{i,k}-E^{(2)}_{j,k} \\
-E_{i}-E_{j}-E_{k} \\
\cdots
\end{gathered}
\end{equation}
$E_{i,j}$ and $E_{i,j,k}$ are energies of dimers and trimers. If the many-body expansion is truncated at second order, the total energy of the system can be approximated as
\begin{equation}
\label{MBE_2}
E_{\text{MBE2}}=\sum^{C^n_2}_{i<j} E_{i,j}-(n-2) \sum^{n}_{i} E_{i}
\end{equation}
If one includes third-order corrections, the total energy is
\begin{equation}
\label{MBE_3}
\begin{aligned}
E_{\text{MBE3}}=& \sum^{C^n_3}_{i<j<k} E_{i,j,k}-(n-3) \sum^{C^n_2}_{i<k} E_{i,j} \\
&+\frac{1}{2}(n-2)(n-3) \sum^{n}_{i} E_{i}
\end{aligned}
\end{equation}
Note that the total energy is exact if no truncation is applied to eq.~\ref{eq:MBE}. When the MBE is truncated at second or third order, additional corrections to the long-range interactions is necessary to recover the exact result. The most popular scheme for MBE corrections is the electrostatic embedding approach, in which the long-range interaction is represented as the point charge interaction.~\cite{DahTru07} To further improve the accuracy of the MBE, it is also possible to use higher level ES methods, such as density function theory or even wave function methods, to describe the long-range interactions. 

%A general version of the MBE (GMBE) has been proposed by Richard and Herbert to enable overlapping fragmentation. Based on the principle of inclusion/exclusion, the total energy and Hamiltonian can be easily defined without overcounting many-body interactions.   

\begin{figure}
    \centering
    \includegraphics[width= 1\linewidth]{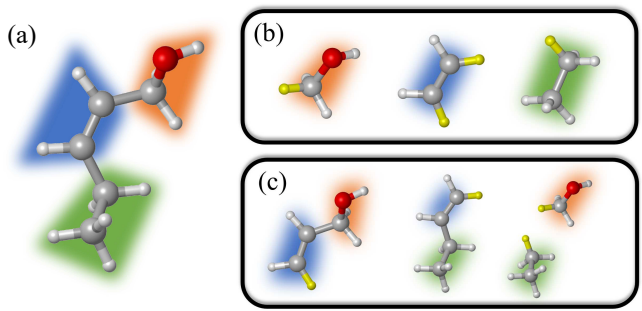}
    \caption{Using (a) Pentenol molecule as an example for identifying (b) monomers and (c) dimers. Hydrogen atoms highlighted in yellow are link atoms added to saturate the severed bonds.}
    \label{fig:link_atom}
\end{figure}

One intrinsic drawback of the fragment-based methods is the broken chemical bonds introduced by the molecular fragmentation. The hydrogen atom is usually chosen to saturate the severed bonds as shown in Figure~\ref{fig:link_atom}. It is important to guarantee that the net number of capped hydrogen atoms is zeros for a valid fragmentation method. This condition is automatically satisfied in the MBE and GMBE methods.

\subsubsection{Density Matrix Embedding Theory}
The DMET method provide another way to evaluate interactions between fragment and environment. In comparison to MBE method, the DMET method does not contain any procedure to cut chemical bonds and deals with embedded Hamiltonians instead. \\

Consider that a large system $\mathcal{S}$ can be decomposed into two subsystems  $\mathcal{S}_\mathrm{A}$ and $\mathcal{S}_\mathrm{B}$, namely $\mathcal{S} = \mathcal{S}_\mathrm{A} + \mathcal{S}_\mathrm{B}$. An arbitrary quantum state $|\Psi\rangle$ of $\mathcal{S}$ can be in general expressed in the Hilbert space of $|\Psi^\mathrm{A}_I\rangle \otimes |\Psi^\mathrm{B}_J\rangle$,
\begin{equation}
|\Psi\rangle =\sum_I^{d_{\mathrm{A}}} \sum_I^{d_\mathrm{B}} C_{IJ} \left|\Psi^\mathrm{A}_{I}\right\rangle \left|\Psi^\mathrm{B}_J\right\rangle, 
\end{equation}
where $\Psi^\mathrm{A}$ ($\Psi^\mathrm{B}$) is the state of $\mathcal{S}_\mathrm{A}$ ($\mathcal{S}_\mathrm{B}$) of dimension $d_{\mathrm{A}}$ ($d_{\mathrm{B}}$). In quantum chemistry simulations, molecular or material systems are often composed of multiple subsystems that are small enough to enable high-level electronic structure calculations.  $\mathcal{S}_\mathrm{A}$ can be considered one of local subsystems and $\mathcal{S}_\mathrm{B}$ is considered the environment that is composed of the rest of subsystems. The central idea of the DMET is to find the Schmidt decomposition of the wave function $|\Psi\rangle$. By the singular value decomposition of the coefficients $\mathbf{C}$, 
\begin{equation}
    C_{IJ} = \sum_{\kappa}^{d_\kappa} U_{I \kappa} \lambda_{\kappa} V_{\kappa J}^{\dagger},
\end{equation}
the Schmidt states can be defined for $\mathcal{S}_\mathrm{A}$ and its DMET bath,
\begin{equation}
    \begin{aligned}
        \left|\tilde{\Psi}^\mathrm{A}_{\kappa}\right\rangle = \sum_{I}^{d_\mathrm{A}} U_{I\kappa} \left|\Psi^\mathrm{A}_{I}\right\rangle \\
        \left|\tilde{\Psi}^\mathrm{B}_{\kappa}\right\rangle = \sum_{J}^{d_\mathrm{B}} V^*_{J\kappa} \left|\Psi^\mathrm{B}_J\right\rangle.
    \end{aligned}
\end{equation}
$d_{\kappa}$ is the rank the coefficient matrix $\mathbf{C}$. Assuming that $\mathbf{C}$ is a full rank matrix, $d_{\kappa}= \mathrm{min}\left(d_\mathrm{A},d_\mathrm{B}\right)$. As such, the electronic structure of the full system can be solved in an embedding representation of the subsystem orbitals plus its bath orbitals 
\begin{equation}
|\Psi\rangle =\sum_{\kappa}^{d_{\kappa}} \lambda_{\kappa} \left|\tilde{\Psi}^\mathrm{A}_{\kappa}\right\rangle \left|\tilde{\Psi}^\mathrm{B}_{\kappa}\right\rangle.
\label{eq:bath_SVD}
\end{equation}
A simple construction of the embedding Hamiltonian in the basis of Schmidt states is 
\begin{equation}\label{eq:embed_H}
\begin{aligned}
    \hat{H}_{\mathrm{emb}} &= \hat{P} \hat{H} \hat{P} \\
    & = \sum_{pq}^{d_\kappa} \tilde{h}_{pq} a_p^\dag a_q + \sum_{pqrs}^{d_\kappa} v_{qs}^{pr} a_p^\dag a_r^\dag a_s a_q 
\end{aligned}
\end{equation}
with the projector
\begin{equation}
\hat{P} =\sum_{\kappa \gamma}\left|\tilde{\Psi}^\mathrm{A}_{\kappa} \tilde{\Psi}^\mathrm{B}_{\gamma}\right\rangle\left\langle\tilde{\Psi}^\mathrm{A}_{\kappa} \tilde{\Psi}^\mathrm{B}_{\gamma}\right|,
\label{eq:DMET_projection}
\end{equation}
and the one-electron coefficient
\begin{equation}\label{eq:embed_h}
    \tilde{h}_{pq} = h_{pq} + \sum_{rs}^N (v_{qs}^{pr} - v_{sq}^{pr}) D_{rs}^{\mathrm{B}}.
\end{equation}
Here, $h_{pq}$ and $v_{qs}^{pr}=(pq|rs)$ are one- and two-electron integrals. Eq.~\eqref{eq:embed_H}-\eqref{eq:embed_h} represent an {\it interaction bath formulation} of the DMET. In contrast, the {\it noninteracting bath formulation} of the embedding Hamiltonian includes only two-electron integrals on the subsystem orbitals and mimics other Coulomb interactions with a correlation potential.~\cite{wouters2016practical}  
%decompose the full system into a large number of small systems, which can be efficiently dealt with high-level quantum mechanics methods. The interaction between the embedded system and the environment is described by including appropriate bath orbitals in the embedding calculation.  

%In the DMET, a mean-field approximation calculation is performed for the full systems. The Schmidt decomposition is carried out for the mean-field wave function, and the electronic structure of the embedded system is solved using a quantum solver. The DMET method provide an efficient way to accurately describe the interaction between the embedded system and the environment by including a quantum bath. 

%However. the ground state of the full system is hardly known, so in practice, a less accurate ground state (often a mean field one) is used, then a accurate VQE calculation is performed to get the ground state and reduced density matrix of the embedded Hamiltonian.

If the wave function of the full system is known a priori, the bath orbitals and the embedding Hamiltonian are well defined. However, due to the high computational complexity for obtaining a high-level wave function of the full system, an approximate wave function from the low-level electronic structure calculation, such as Hartree-Fock, is used to construct the bath orbitals. These bath orbitals can be self-consistently improved by solving the embedded problems with a high-level electronic structure method. Given the wave function approximation introduced in the DMET, the embedding density matrix for the subsystem $S_\mathrm{A}$ may differ from the exact one. To ensure that the total number of electron in the subsystems adds up to $N_e^\mathrm{A}$, a global potential $\mu$ is applied to fix the embedded Hamiltonian:
\begin{equation}
\hat{H}_{new}^{\mathrm{emb}} \leftarrow \hat{H}_{old}^{\mathrm{emb}}-\mu \sum_{p \in \mathrm{A}} \hat{a}_{p}^{\dagger} \hat{a}_{p}
\end{equation}

%Further more a cost function is needed to evaluate the general potentials used. As is mentioned before, the mean-field ground state will cause differences in density matrix, so the variance of the density matrix is a good cost function to use:
%\begin{equation}
%\mathcal{L}=\sum_{x} \sum_{r s \in Frag_{x}}\left(D_{r s}^{x}-D_{r s}^{\mathrm{mf}}\right)^{2}
%\end{equation}

In Ref.~\citenum{wouters2016practical}, there are some of cost functions summarized to optimize the low-level Hamiltonian and correlation potential, including fragment plus bath fitting, fragment only fitting, and single-shot embedding. In this work, the simplest cost function 
\begin{equation}
\mathcal{L}=\left(\sum_{A,p} D_{pp}^{A} - N_{\text {e }}\right)^{2}
\end{equation}
and the fragment only fitting scheme 
\begin{equation}
\mathcal{L}=\sum_{A} \sum_{r s \in A}\left(D_{r s}^{A}-D_{r s}^{\mathrm{mf}}\right)^{2}
\end{equation}
has been employed to optimize the correlation potential.

\subsection{Adaptive VQE Algorithms}
The accuracy and circuit depth of the VQE method depend heavily on the wave function ansatz. In contrast to the UCC ansatz, the recently proposed ADAPT ansatz\cite{ADAPT_VQE_1} provides an exact quantum solver of electronic structure problems. Meanwhile, the ADAPT ansatz bypasses the Trotterization error of the UCC ansatz by generating a pseudo-Trotter wave function with a compact sequence of unitary transformations acting on the reference state
	\begin{equation}
		|\Psi^{\mathrm{ADAPT}}_{(k)} \rangle =(e^{\theta_k} \hat{\tau}_k) \cdots (e^{\theta_{1} \hat{\tau}_{1}})|\Psi_{0} \rangle
	\end{equation}
where anti-Hermitian operators $\hat{\tau_{i}} \in \mathcal{O} \equiv \{\hat{\tau}_{q}^{p}, \hat{\tau}_{rs}^{pq}\}$:
	\begin{equation}
		\hat{\tau}_{q}^{p}=a_{p}^{\dagger}a_{q}-a_{q}^{\dagger}a_{p}
	\end{equation}
	\begin{equation}
		\hat{\tau}_{rs}^{pq}=a_{p}^{\dagger}a_{q}^{\dagger}a_{r}a_{s}-a_{s}^{\dagger}a_{r}^{\dagger}a_{q}a_{p}
	\end{equation}
At the $k$th iteration, the energy is minimized through a conventional VQE procedure:
	\begin{equation}
		E^{(k)}=\min_{\vec{\theta}^{(k)}}{\{\langle \Psi_\text{ADAPT}^{(k)}(\vec{\theta}^{(k)})|\hat{H}|\Psi_\text{ADAPT}^{(k)}(\vec{\theta}^{(k)}) \rangle \}}
	\end{equation}
The residual gradient of the $i$th operator $\hat{\tau}_{i}$ in $\mathcal{O}$ must evaluated at each iteration to determine the sequence of the pseudo-Trotter ansatz:
	\begin{equation}
		\begin{aligned}
			R_{i}^{(k)}&=\langle \Psi_\text{ADAPT}^{(k)} |[\hat{H},\hat{\tau}_{i}]| \Psi_\text{ADAPT}^{(k)}  \rangle \\
			&=\left.\frac{\partial{E^{(k+1)}}}{\partial{\theta_{k+1}^{(k+1)}}}\right|_{\theta_{k+1}^{(k+1)}=0,\hat{\tau}_{k+1}^{(k+1)}=\hat{\tau}_{i}}
		\end{aligned}
	\end{equation}
The operator $\hat{\tau}^{(k+1)}$ with the largest \textit{pre-estimated} gradient $R^{(k)}$ is selected from $\mathcal{O}$ and added at the next iteration to update the wave function ansatz
	\begin{equation}
		|\Psi_\text{ADAPT}^{(k+1)}=e^{\theta_{k+1}^{(k+1)}\hat{\tau}^{(k+1)}} | \Psi_\text{ADAPT}^{(k)} \rangle
	\end{equation}
The convergence criteria can be set by tracking the norm of \textit{pre-estimated} gradients $\vec{R}_{k}=(R_{k}^{N_{k}},R_{k}^{N_{k}-1},\dots,R_{k}^{1})$:
	\begin{equation}
		\label{eq-adapt-convg}
		\Vert\vec{R}^{(k)} \Vert_{2}=\sqrt{\sum_{i} \left(R_{i}^{(k)}\right)^2}<\varepsilon
	\end{equation}
	
The fermionic creation and annihilation operators are mapped onto qubit operators $\{X, Y, Z, I\}$ using transformations, such as the Jordan-Wigner\cite{JW_1,JW_2} or Bravyi-Kitaev\cite{BK_1,BK_2,BK_3}. The operator pool in adaptive VQE algorithms can be constructed based on a variety of operator forms. In addition to fermionic operators , another widely used form of excitation operators are qubit operators. As introduced in the qubit-ADAPT-VQE ansatz\cite{qubit-adapt-vqe}, the Pauli string operators with a maximal length of 4 are used to reduce the number of C-NOT gates in the quantum circuit. Since the qubit operators only contain Pauli strings, they are promising in building hardware-efficient circuits at the same computational accuracy. One scheme to construct qubit operators is to choose individual Pauli strings after the fermionic operators in the unitary coupled cluster ansatz are mapped onto qubit operators. The qubit operator generally have the form: 
\begin{equation}
    \hat{\tau}=i \prod_{i} p_{i}, \quad p_{i} \in \{X, Y, Z\}.
\end{equation}
Another one is qubit excitation based operator (QEB)~\cite{QEB} in the form of constructed by qubit creation and qubit annihilation operators. In qubit-ADAPT-VQE and QEB, Pauli $Z$ strings are discarded in order to construct shallow circuits. As a consequence, more circuit parameters may be required to restore the antisymmetry of the wave function.

\subsection{Implementation}
Here, we briefly summarize the details of implementing the MBE-VQE and DMET-VQE methods.
\begin{figure}
    \centering
    \includegraphics[width = \linewidth]{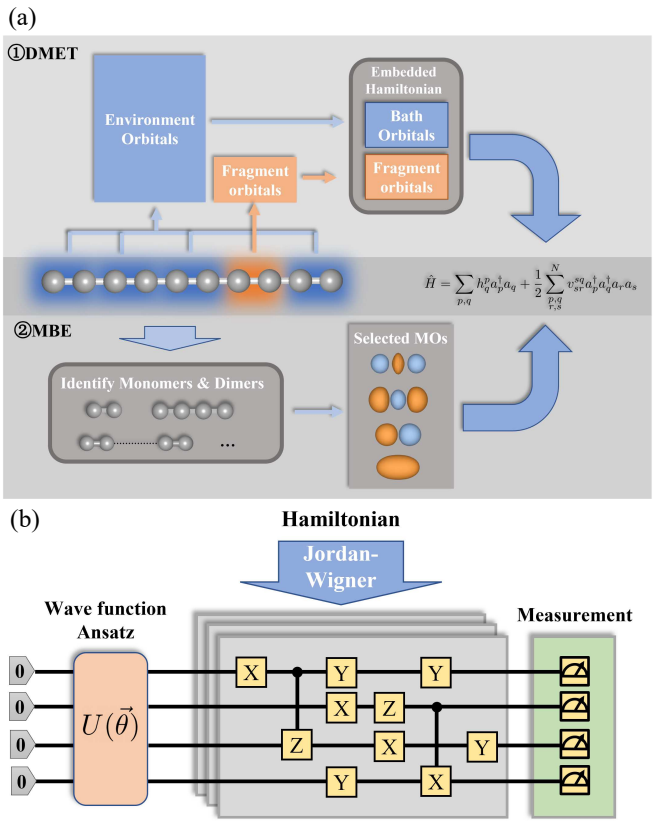}
    \caption{(a) Brief workflow of how to get the Hamiltonian in DC methods. (b) Schematic of quantum circuit used to solve the ground state of the Hamiltonian.}
    \label{fig:implementation}
\end{figure}
\subsubsection{MBE-VQE}
The implementation of the MBE-VQE method can be simply described as
\begin{itemize}
    \item Divide the full system into fragments with appropriate sizes.
    \item Determine $n$-mers required to compute the many-body corrections.
    \item Add link atoms at proper positions if necessary.
    \item Perform VQE calculations for monomers, dimers, trimers, and so forth.
    \item Calculate the total energy following the MBE energy expression.
\end{itemize}
In this work, the C-C bonds are severed when fragments are defined for C$_{18}$. A hydrogen atom is places at a position $\mathbf{r}_\text{cap}$ between two Carbon atoms at $\mathbf{r}_1$ and $\mathbf{r}_2$. The direction of the C-H bond is the same as that of the original C-C bond, and its bond length is set to 1.061. The coordinate of $\mathbf{r}_\text{cap}$ is 
\begin{equation}
    \mathbf{r}_\text{cap} = \mathbf{r}_\text{1}+\frac{\mathbf{r}_\text{2}-\mathbf{r}_\text{1}}{|\mathbf{r}_\text{2}-\mathbf{r}_\text{1}|} \mathbf{r}_{\text{C-H}}.
\end{equation} 

In comparison with the DMET-VQE method, one of the most important merits of the MBE-VQE method is independent calculations for subsystems with massive parallel. And these parallel does not require extensive modification to existing VQE algorithms. In addition, the long-range interaction can be restored with low level electronic structure methods, which often requires little computational cost. However, it is necessary to avoid severing multiple bonds in the MBE-VQE method because it is not a trivial work to saturate these severed bonds. 

\subsubsection{DMET-VQE}
The DMET-VQE method uses a VQE solver to determine the ground state wave functions of the embedded Hamiltonians constructed by the DMET algorithm. Note that the Hamiltonian contain both fragment orbitals and bath orbitals, so the reduced density matrices (1-RDM and 2-RDM) are computed to get the expectation values (namely the electron number and energy) of the fragment. The total electron number and total energy of the whole system is then obtained by adding up electron numbers and energies of all fragments.

The pseudo-code of the DMET-VQE method is shown below:\\
\begin{algorithm}[H]
\caption{Pseudo-code for DMET}
\LinesNumbered %要求显示行号
\KwIn{Low-level wave function of the full system $ |\Psi_0\rangle$}%输入参数
\KwOut{The total energy and reduced density matrices}%输出
\While{$|\mathcal{L}|>\tau$}{
    \For{fragments in molecule}{
    Construct bath orbitals $|B\rangle$ $\leftarrow |\Psi_0\rangle$ \;
    Build the embedding Hamiltonian $H_{\rm emb} \leftarrow |\Psi_0\rangle,|B\rangle $ \;
    Diagonalize the embedding Hamiltonian with the VQE algorithm :$|\Psi\rangle$\ $\leftarrow H_{\rm emb}$ \;
    Compute \text{1-RDM} and \text{2-RDM} $\leftarrow |\Psi\rangle $ \;
    Compute the energy ($E_{\rm frag}$) and the number of electrons ($N_{\rm frag}$) of the fragment $\leftarrow$ \text{1-RDM} and \text{2-RDM}\;
    }
    $N_{\rm elect}=\sum_{\rm frag}N_{\rm frag}$\;
    $E_{\rm total} = \sum_{\rm frag}E_{\rm frag}$\;
    Compute the cost function $\mathcal{L}$
}
\end{algorithm}
In this work, we use Hartree-Fock as the low-level method to compute the wave function of the full system. The meta L\"owdin localization scheme is used to construct localized orbitals.

\subsubsection{Dependencies}
The PySCF package \cite{sun2018pyscf,sun2020recent} is used for restricted and unrestricted Hartree-Fock and CCSD calculations.  The one- and two-electron integrals for constructing the second quantization Hamiltonian in VQE calculations are also extracted from PySCF. The DMET calculations have been performed with QC-DMET\cite{wouters2016practical}. The complete active space configuration interaction (CASCI) calculations are carried out with the CheMPS2\cite{wouters2014chemps2} package. An in-house VQE simulation package~(Q$^{2}$Chemistry
\cite{fan2022q}
) is used for the VQE calculations.  %OpenFermion~\cite{mcclean2020openfermion} is used to carry out Jordan-Winger transformation and generate sparse matrices. 

\section{Results}
\label{sec:result}  

\subsection{Hydrogen Chain}

\begin{figure}
    \begin{center}
    \includegraphics[width=1\linewidth]{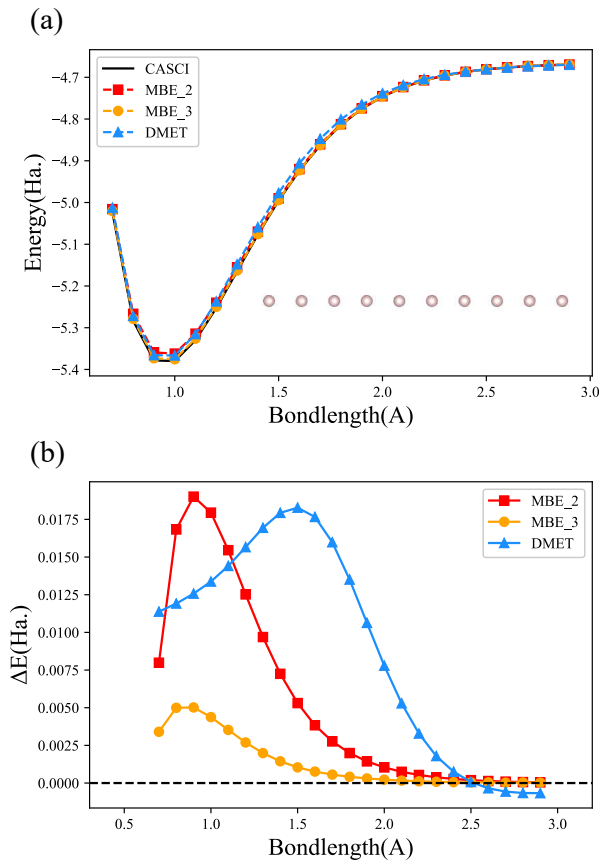}
    \end{center}
    \caption{Potential energy surfaces computed with MBE-VQE and DMET-VQE for the hydrogen chain with 10 hydrogen atoms equispaced along a line and errors. The reference values are the CASCI results. All numerical simulations are performed with the STO-3G basis.
	}
    \label{fig:01_H10_energy}
\end{figure}
%\begin{figure}
%    \centering
%    \includegraphics[width=1\linewidth]{Be_jupyter.eps}
%    \caption{The simulated energy curve of beryllium ring with %   DMET-VQE, in comparison with DMET-FCI and CCSD. }
%    \label{fig:02_Be_energy}
%\end{figure}
We first compute the potential energy surface (PES) of the hydrogen chain with 10 hydrogen atoms equispaced along a line using the MBE-VQE and the DMET-VQE methods. The hydrogen chain is a extremely simple model while it is essential for understanding diverse fundamental physical phenomena, such as insulator-to-metal transition and antiferromagnetic Mott phase.~\cite{MotGenMa20} Here, the hydrogen chain is divided into five fragments and each fragment contains 2 hydrogen atoms. Two kinds of MBE-VQE methods, including MBE2-VQE and MBE3-VQE, have been employed to describe high-order corrections. In MBE2-VQE, dimers consisting of 4 hydrogen atoms are able to compute second-order corrections. Analogously, trimers consisting of 6 hydrogen atoms in MBE3-VQE are used to compute third-order corrections. All calculations are performed with the STO-3G basis. The maximal number of qubits used in MBE2-VQE, MBE3-VQE and DMET-VQE is 8, 12 and 8, respectively. 

Figure.~\ref{fig:01_H10_energy} shows PESs and their error curves with respect to the CASCI results for H$_{10}$. The maximal deviations in MBE2-VQE and DMET-VQE are almost the same. The MBE2-VQE method and the DMET-VQE method use the same number of qubits, but the MBE2-VQE method has a lower average deviation (5.46 millihartree for MBE2-VQE and 9.00 millihartree for DMET-VQE). As shown in Figure ~\ref{fig:01_H10_energy}, it is clear that, as the bond length increase, the total energy deviations decrease in all three methods. This is consistent with the intuition that the DC methods are more accurate when the interactions between fragments become smaller. When the hydrogen atoms are well separated, namely the H-H bond length larger than 2.5 \AA, the errors in both MBE2-VQE and DMET-VQE are less than 1 millihartree. In comparison with MBE2-VQE, the overall errors of MBE3-VQE is much smaller. For MBE2-VQE and MBE3-VQE, the errors near the equivalent bond length are larger than those with elongated H-H bond length.

\begin{figure}
	\centering
	\includegraphics[width=1\linewidth]{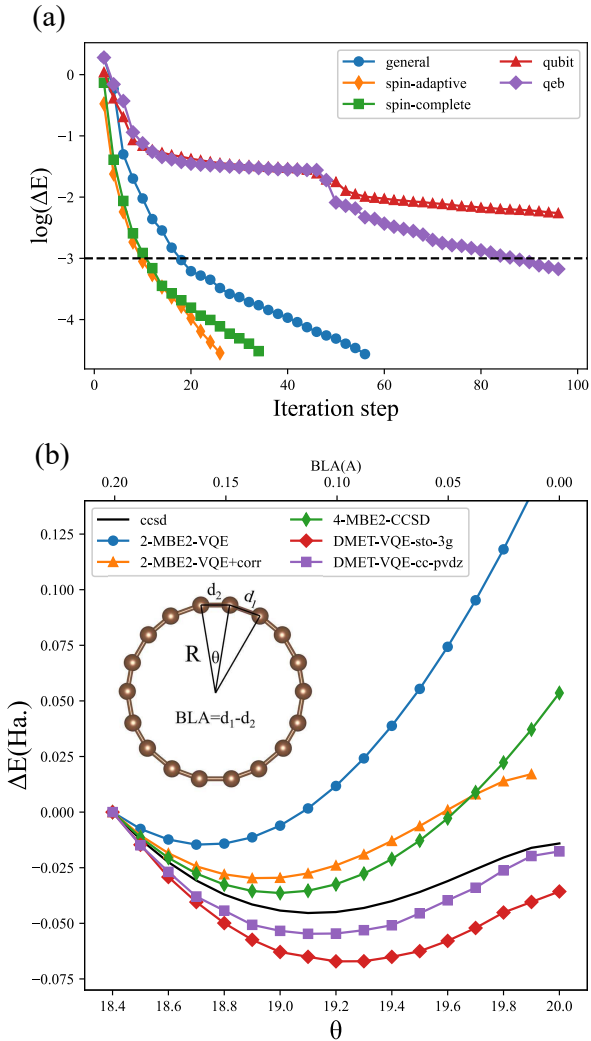}
	\caption{(a) Errors of the total energy as a function of the number of iterations. Three types of fermionic operators, (general, spin-complete and spin-adaptive anti-Hermitian operators) 
	and two types of qubit operators, including multi-qubit and qubit excitation based (QEB) operators, are tested. (b) Relative energy $(\Delta E=E-E_{\theta = 18.4})$} versus $\theta$ and bond length alternation(BLA) of various methods. n-MBE2 means a MBE2 calculation while n carbon atoms are considered as a fragment.
	\label{fig_C18}
\end{figure}

%\begin{figure}
%	\centering
%	\includegraphics[width=1\linewidth]{6_H2O_all_new.eps}
%	\caption{relative energy with different $d_1-d_2$ values with various methods. The three structures above shows the start, intermediate, and final structure of the proton transfer process}
%	\label{6H2O}
%\end{figure}

\subsection{Cyclo[18]carbon}
%\change{Discussion about ansatz? no data is presented} 
The operators used in the adaptive VQE algorithm can be generally divided into two types, namely fermionic operators and qubit operators. Qubit operators are expected to be circuit-depth friendly while fermionic operators can consider many symmetries of the wave function, such as antisymmetry and spin symmetry, which make them easy to achieve the convergence of the electronic energy. Here, we investigate the convergence of different operators for the embedded Hamiltonian constructed in the DMET method with a cyclo[18]carbon molecule. Three types of fermionic operators, including general, spin-complete and spin-adaptive anti-Hermitian operators, together with two types of qubit operators, including multi-qubit and QEB operators, are tested. The convergence criteria is consistently set to $\left\langle\psi\left|H^{2}\right| \psi\right\rangle-\langle\psi|H| \psi\rangle^{2} \leq 0.01$. As is shown in figure\ref{fig_C18}(a), calculations with fermionic operators reach convergence within 30 iterations while qubit operators fail to do so with a much larger number of iterations. In particular, the QEB operators were demonstrated to be able to achieve the same rate of convergence as the general fermionic operators in previous studies.~\cite{QEB,LiuLiYan21} While, in the DMET-VQE calculations, the convergence of the QEB operators is much slower than the general fermionic operators. All calculations in the rest of this work use spin-adapted fermionic operators to construct operator pools for adaptive VQE algorithms.

Cyclo[n]carbons have attracted much attention as a molecular carbon allotrope consisting two-coordinated carbon atoms.~\cite{DieRubKno89,ParAlmFey91,HelGotBow93,HutLueDie94} Recently, the cyclo[18]carbon molecule have been successfully synthesized in solid state and its structure has been characterized by high-resolution atomic force microscopy and density functional theory.~\cite{C18} Here, Density functional theory calculations using the HSE hybrid functional, reveal that the cyclo[18]carbon exhibits a $D_{9h}$ symmetry and the short carbon-carbon (C-C) bond length is $\SI{1.195}{\angstrom}$ and the long C-C bond length is $\SI{1.343}{\angstrom}$. This result is consistent with the coupled cluster calculation that estimates the bond length alternation (BLA) to be $\sim$ 0.14 \AA.~\cite{StaStaSol20} The BLA is defined as the difference between the long and short C-C bond length.

In this work, we apply the MBE-VQE and DMET-VQE methods to study relative energies of the cyclo[18]carbon ring. A series of cyclo[18]carbon structures are generated by fixing the diameter of the ring to \SI{7.31}{\angstrom} while changing the BLA. Figure~\ref{fig_C18}(b) shows the relative energies as a function of the BLA. Only important orbitals, namely 2s and 2p$^3$ orbitals of carbon atoms, are considered in VQE calculations with the cc-pVDZ basis and other orbitals are treated at a mean-field level. The DMET-VQE calculations with both the sto-3g and cc-pVDZ basis are highly consistent with the CCSD result with the cc-pVDZ basis. However, the DMET-VQE calculations indicates that the structure with the lowest energy has a BLA of \SI{0.11}{angstrom}, which is slightly smaller then that obtained in ref.~\citenum{C18}. In the MBE-VQE method, it is inappropriate to take one carbon atom as one fragment because the triple bonds will be served in this situation. As such, we divide cyclo[18]carbon into fragments with 2 and 4 atoms. We label the second order MBE calculation with n carbon atoms in a fragment as n-MBE2. In the 2-MBE2-VQE calculation, the STO-3G basis is used and eight molecular orbitals are considered as active orbitals for dimers and four molecular orbitals are considered active for monomers. A dimer in 4-MBE2 consists of 8 carbon atoms. Dealing with such a large fragment exceeds the limit of our quantum simulator's capabilities, so that the ground state energy is computed with the CCSD method with cc-pVDZ basis. Figure~\ref{fig_C18} shows that ,as BLA approaches 0, the deviation caused by MBE keeps getting larger. This is because, when BLA gets smaller, the interactions between fragments get stronger and the boundary of fragments become ambiguous. In such case, the severed C-C bonds can be not simply saturated by the hydrogen atoms. Instead, ghost orbitals introduced in FMO method~\cite{kitaura1999fragment,nakano2000fragment,kitaura2001fragment} or the overlapping fragments introduced in the generalzied MBE approached~\cite{RicHer12} should be used to correct errors induced by severing C-C bonds. To restore the long-range interactions, we introduce a RHF correction to the MBE2 approach,that is, 
\begin{equation}
\begin{aligned}
    E_{\rm MBE2-VQE+corr}&=E_{\rm MBE2-VQE}+E_{\rm corr}\\
    E_{\rm corr}=&E_{\rm RHF}-E_{\rm MBE2-RHF}.
\end{aligned}
\end{equation}
%\begin{equation}
%    E_{MBE2-VQE+corr}=E_{MBE2-VQE}+(E_{RHF}-E_{MBE2-RHF}).
%\end{equation}
 Although the deviations of the MBE2-VQE method gets large when the BLA become small, it still gives a good estimate of the equivalent bond length. The MBE methods indicate the lowest energy at a BLA of \SI{0.138}{\angstrom} and \SI{0.127} with 2-MBE2-VQE plus the RHF correction and 4-MBE2-CCSD, respectively. 

\begin{figure}
    \centering
    \includegraphics[width = \linewidth]{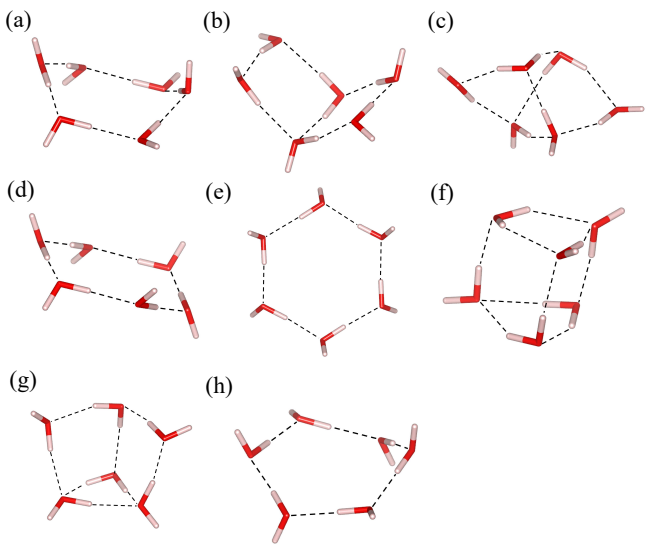}
    \caption{A series of typical structures of water hexamers, (a)boat (b)book (c)cage (d)chair (e)cyclic (f)prism (g)prism-book(p-b for short) (h)twisted-boat (t-boat for short)}
    \label{fig:water_hexamer_structure}
\end{figure}

\begin{figure}
    \centering
    \includegraphics[width = \linewidth]{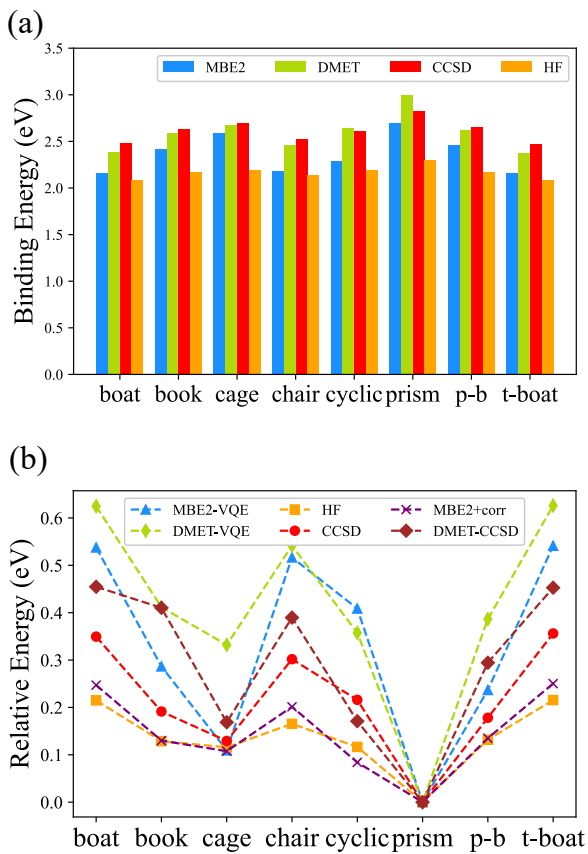}
    \caption{(a)Binding energies and (b)Relative energies (in eV) of different water hexamer structures computed with Hartree-Fock, DMET-VQE, MBE2-VQE, and CCSD methods. The energy of the prism hexamer is used as a reference value.}
    \label{fig:water_hexamer_binding_energy}
\end{figure}

\subsection{Water Hexamers}
As the most important substance for life, various properties of water have been widely studied.
As the smallest piece of the ice, the structure of water-hexamer, namely the cluster composed of six water molecules, have attracted much attention\cite{bilalbegovic2010water_hexamer,hincapie2010water_hexamer,kim1994water_hexamer,nauta2000formation} because it helps us to understand the behavior of water. It is interesting to apply divide-and-conquer methods to such systems. 

Figure~\ref{fig:water_hexamer_structure} shows eight of the most stable water hexamer structures determined in ref.~\citenum{hincapie2010water_hexamer}. An annealing strategy is used to find a series of stable conformations at the MP2/6-311++g** level. The most stable structure is estimated to be the prism one. The total energies of these water hexamers are computed with Hartree Fock, DMET-VQE, MBE2-VQE and CCSD method. In the DMET-VQE and MBE2-VQE method, each water molecule is specified as one fragment. The basis set is cc-pVDZ. To construct DMET bath orbitals, four orbitals (2s,2p$^3$ orbitals of oxygen atom) are considered as the subsystem orbitals, and other orbitals in the fragment together with orbitals in other fragments are compressed into four bath orbitals. As such, the number of qubits in the DMET-VQE method is 16. MBE-VQE calculation of the water hexamer requires ground energy of water dimers. To reduce the computing resource, eight molecular orbitals closest to HOMO and LUMO are taken as active orbitals.\\

Figure~\ref{fig:water_hexamer_binding_energy} shows binding energies and relative energies of eight different structures of water hexamers. The binding energy is defined as the total energy of hexamers minus the energy summing up energies of six isolated water molecules. Relative energy is defined as the relative values with respect to the  energy of the most stable hexamer, namely the prism hexamer. Note that all four methods give a correct prediction of the most stable structure. In comparison with other three methods, the DMET method shows relatively large deviations. There are two main reasons for such deviations: one is the systematic error of the DMET method; another is a small number of orbitals used in the DMET method when we construct bath orbitals. We perform the DMET-CCSD calculations with a larger active space in the fragment (1s,2s,3s,$2p^3$,$3p^3$ orbitals for oxygen atoms and 1s,2s orbitals for hydrogen atoms), which gives a more reliable result. It is worth mentioning that, although the binding energies from MBE-VQE and DMET-VQE methods are highly consistent with the CCSD results in some conformations, including prism, cage and prism-book, there are innegligible deviations in the rest of eight conformations. This mainly results from the DC strategy. Especially, in the MBE-VQE method, some interactions in the system are completely ignored. This is a systematic error of the MBE-VQE method. Fortunately, this systematic error (i.e. $E_\text{VQE}-E_\text{MBE-VQE}$) is roughly equivalent to $E_\text{RHF}-E_\text{MBE-RHF}$. Figure ~\ref{fig:water_hexamer_binding_energy} shows that the MBE plus such a correction gives a better result.

%\subsection{Ligand: Chloroquine}

%\begin{figure}
%\centering
%  \includegraphics[width=0.6\linewidth]{06_Chlorogquine.PNG}
%  \caption{The structure of Chloroquine.   }
%  \label{fig:06_Chloroquine}
%\end{figure}

%\begin{table}[h]
%\centering
%\begin{tabular}{|l|c|}\hline
%Method&Energy (Ha.)\\\hline
%RHF&-1315.9039 \\
%CCSD&-1317.3365 \\
%DMET-CCSD&-1317.6390 \\
%DMET-DMRG&-1317.6381 \\
%DMET-VQE & -1317.58313\\\hline
%\end{tabular}
%\caption{Calculated energies of Chloroquine with various methods}
%\label{tab:Margin_settings}
%\end{table}

\section{Conclusions}
\label{sec:conclusions}
In this work, we integrate two different divide-and-conquer schemes, including the many-body expansion fragmentation approach and the density matrix embedding approach, into the adaptive VQE algorithms. These DC schemes can be easily implemented within the VQE framework without extensive modification to existing quantum algorithms. The MBE fragmentation approach is more suitable for describing the strongly correlated intra-fragment interactions, leaving the weak inter-fragment interactions to the high-order many-body corrections. This implies that in principle multiple chemical bonds should be not severed in the fragmentation. The calculations for monomers, dimers, trimers, ... in the MEB fragmentation approach are highly independent so that the MBE-VQE approach is embarrassingly parallelizable if a large number of quantum processors are available. In addition, the MBE-VQE approach can be easily combined with posterior correction methods to restore the long-range interactions when the many-body expansion is truncated at a low order. The DMET-VQE method is a more flexible DC scheme, which can be applied to general situations. In order to improve the accuracy of the DMET-VQE method, one can increase the size of fragments and use a more accurate electronic structure method to replace the mean-field approximation.  

The MBE-VQE and DMET-VQE approaches have been applied to study the cyclo[18]carbon molecule and water hexamers. The MBE-VQE approach gives a good description of relative energies of these two systems. For the cyclo[18]carbon molecule, the equivalent bond length predicted by the MBE-VQE approach agrees well with the existing results. For water hexamers, the MBE-VQE approach gives a good estimate of the relative stability of different configurations. The combination of DC methods with more efficient quantum simulators can significantly increase these methods' capability. ref.\citenum{shang2022large} uses a MPS simulator on the new sunway supercomputer, achieved incredible efficiency for quantum chemistry simulation. The DC methods are expected to find promising applications in the fields like biochemistry, catalysis or other complicated chemical reactions.

\section{Acknowledgments}
This work was supported by Innovation Program for Quantum Science and Technology (2021ZD0303306), the National Natural Science Foundation of China (22073086, 22003073, T2222026, 21825302 and 22288201), Anhui Initiative in Quantum Information Technologies (AHY090400), and the Fundamental Research Funds for the Central Universities (WK2060000018).

\section{Data Availability}

The data that supports the findings of this study are available within the article and its Appendix.

%%%%%%%%%%%%%%%%%%%%%%%%%%%%%%%%%%%%%%%%%%%%%%%%%%%%%%%%%%%%%%%%%%
%                        Bibliography                            %
%%%%%%%%%%%%%%%%%%%%%%%%%%%%%%%%%%%%%%%%%%%%%%%%%%%%%%%%%%%%%%%%%%
\bibliography{manuscript} % Produces the bibliography via BibTeX.
\end{document}